\documentclass[manuscript]{aastex}
\usepackage{natbib}
\usepackage{amsmath}
\usepackage{multirow}

\shorttitle{Solar Limb}
\shortauthors{De la Luz, V.}

\begin{document}

\title{The Chromospheric Solar Limb Brightening at Radio, Millimeter, Sub-millimeter, and Infrared Wavelengths}


\author{V. De la Luz}
\affil{CONACYT - SCiESMEX, Instituto de Geof\'isica, Unidad Michoac\'an, Universidad
Nacional Aut\'onoma de M\'exico, Morelia, Michoac\'an, M\'exico. 58190.}

%
%
%

%

\begin{abstract}

  Observations of the emission at radio, millimeter, sub-millimeter, and infrared wavelengths in the center of the solar disk validate the auto-consistence of semi-empirical models of the chromosphere. Theoretically, these models must reproduce the emission at the solar limb. 
  In this work, we tested both the VALC and the C7 semi-empirical models by computing their emission spectrum in the frequency range from 2 GHz to 10 THz, at solar limb altitudes. 
  We calculate the Sun's theoretical radii as well as their limb brightening. 
    Non-Local Thermodynamic Equilibrium (NLTE) was computed for hydrogen, electron density, and H-. In order to solve the radiative transfer equation a 3D geometry was employed to determine the ray paths and Bremsstrahlung, H-, and inverse Bremsstrahlung opacity sources were integrated in the optical depth. We compared the computed solar radii with high resolution observations at the limb obtained by Clark (1994). 
We found that there are differences between observed and computed solar radii of $12000$ km at $20$ GHz, $5000$ km at $100$ GHz, and $1000$ km at $3$ THz for both semi-empirical models. A difference of $8000$ km in the solar radii was found comparing our results against heights obtained from H$\alpha$ observations of spicules-off at the solar limb. We conclude that 
  the solar radii can not be reproduced by VALC and C7 semi-empirical models at radio - infrared wavelengths. Therefore, the structures in the high chromosphere provides a better measurement of the solar radii and their limb brightening as shown in previous investigations.

\end{abstract}






\section{Introduction}
        Following the classical theory of stellar atmospheres, the outer layers of the solar atmosphere must present a single gradient of temperature \citep{1983psen.book.....C} from the photosphere to the interplanetarium medium. However, observations of the quiet Sun from EUV to Radio in the center of the solar disk \citep{1949AuSRA...2..198P,1953sun..book..207V,1991ApJ...370..779Z,2007ApJ...664.1214P,2010SoPh..261...53V} show that in order to reproduce the spectra, it is necessary a complex atmosphere structure. Even more, observations by \cite{1994Sci...263...64S} confirms the existence of a cool region in the chromosphere, where the CO molecula limit the temperature up to $4000$ K. Therefore, the CO molecula fix a lower temperature threshold in the chromospheric models and is called ``the Temperature Minimum of the Sun''.
          
        The chromospheric models include hydrostatic \citep{1953sun..book..207V, 1963IAUS...16....1A,1978SvA....22..345K,1981SoPh...69..273A,1981ApJS...45..635V,2003ApJ...589.1054L,2004A&A...419..747L,2004A&A...422..331C,2008A&A...480..839F}, hydrodynamic \citep[HD,][]{1995ApJ...440L..29C, 1997ApJ...481..500C,2002ApJ...572..626C}, and magnetohydrodynamic (MHD) approximations \citep{2015A&A...575A..15L}. However, the dynamics of the dominant force that allow the existence of CO emission remains as an open question \citep{2014LRSP...11....2P}.

        Despite the simplifications of the hydrostatic atmospheres, the semi-empirical models still useful to compute the flux from solar-like atmospheres \citep{2015A&A...573L...4L} and flares events \citep{2015SoPh..290.2809T} at radio - infrared wavelengths.

        The semi-empirical models with hydrostatic aproximation are focused in reproduce the emission in the center of the solar disk. Theoretically, these models must reproduce the emission at the solar limb but analysis in these regions are not included in the atmosphere computations. One of the most important characteristic in these upper region of the solar atmosphere is that unlike the limb darkening at visible wavelengths \citep{1946ApJ...104...60K}, there is a limb brightening at radio frequencies \citep{1947RSPSA.190..357M}
contributing to an
increase in the apparent solar radius \citep{1950AuSRA...3...34S}.

Earlier observations between 5 GHz (6 cm) and 33 GHz (9 mm) show clearly a limb brightening \citep{1979ApJ...234.1122K}. Observations at shorter 
wavelengths 
(33 GHz and 86 GHz) reported both limb darkening and sharp cutoff 
distribution \citep[non-limb darkening,][]{1972A&A....21..119L,1974ApJ...187..389A}. However, observations with the James Clerk Maxwell radio telescope clearly show solar limb brightening at 850 GHz, 353 GHz, and 250 GHz \citep[350, 850, and 1200 $\mu$m respectively,][]{1995ApJ...453..511L}.  

First attempts to found the sources of these higher emissions between 0.01 GHz (30 m) and 30 GHz (1 cm) at the solar limb showed that the main contributors are the chromosphere and the corona \citep{1946Natur.158..632M, 1946ApJ...104...60K,1947Natur.159..506S,1948Natur.161..133G}. 
The role of the fine-structure involved in the quiet Sun emission (spicules) and its relation with the limb brightening at millimeter wavelengths was discussed in the firsts { semi-empirical models} \citep{1979SoPh...63..257F, 1981SoPh...69..273A}. Further investigations made by \cite{2005A&A...440..367S} showed that the spicules could modulate the morphology of the limb brightening profile.
However, the chromospheric empirical model used by \cite{2005A&A...440..367S} is not necessarily a good approximation for the physical conditions in the {\bf low} chromosphere \citep{2002ApJ...572..626C, 2011ApJ...737....1D}
when considering a full ionized chromosphere.
Observations of the solar limb at H$\alpha$ \citep{2014ApJ...795L..23S} clearly show spicules in the high chromosphere.
Furthermore, \cite{1995hola} found evidence of solar limb occultation in the radio emission from eruptive events located around the limb.


In this work, we applied the numerical code PakalMPI \citep{2011ApJ...737....1D} to solve the radiative transfer equation \citep{2010ApJS..188..437D} to compute the theoretical spectrum from 2 GHz to 10 THz at limb altitudes using as input VALC \citep{1981ApJS...45..635V} and C7 \citep{2008ApJS..175..229A} semi-empirical models. The computed synthetic spectrums are compared against observations by \cite{1994IAUS..154..139C} to test the autoconsistence of VALC and C07 models at the limb. We applied the 3D geometry of PakalMPI to calculate the local emission and absorption processes at several altitudes above the solar limb using the following:
i) three opacity sources: Bremsstrahlung, H${}^-$, and inverse Bremsstrahlung, ii) deriving the maximum relative limb brightening from the chromospheric contribution detailing the local emission and absorption process, iii) a numerical approximation to estimate the solar radii for the frequencies observed, and iv) an estimate of the changes in the solar radii

In Section 2, we introduce the chromospheric model. In Section 3, we show the opacity sources and the theoretical computations for the simulated spectra. In Section 4,
the results of the comparison of our synthetic spectra for VALC and C7 models versus the observations at millimeter - infrared wavelengths in the solar limb are given. In
Section 5, we present our conclusions.

\section{The Chromospheric Model}
In a previous paper \citep{2011ApJ...737....1D}, we introduced the Non-Local Thermodynamic Equilibrium (NLTE) computation of the simulated spectra using an interpolation of the pre-computed departure coefficient $b_1$ \citep{1937ApJ....85..330M} { for hydrogen. The b1 parameter is defined as $$b_1=\frac{n_1/n^*_1}{n_k/n^*_k},$$ where $n^*_1$ and $n^*_k$ are the densities of hydrogen in the ground state $n=1$ and ionized hydrogen in thermodynamic equilibrium (LTE), respectively. The $n_1$ and $n_k$ represents the same but in NLTE. The b1 parameter shows if the system is in LTE ($b_1=1$) or NLTE ($b_1 \ne 1$).}
With this technique, we improved the computation
time and reduced the complexity of the numerical solution. We applied b1 parameters published in \cite{1981ApJS...45..635V}. 
%
The input models are: the VALC from \cite{1981ApJS...45..635V} and the C7 from \cite{2008ApJS..175..229A}. We used these models to compare if the inclusion of the ambipolar diffusion in the C7 chromospheric model is different, if at all, with the classic VALC model in the emission at the limb.

We assumed a static corona as boundary condition for the numerical model. The inclusion of the corona does not modify the final brightness temperature ($T_{\mathrm{b}}(\nu)$) in the frequencies ($\nu$) under study.  

In the VALC model (black lines in Figure \ref{temperature-corona-VALC-C07.eps}), we used observations from \cite{1998A&A...336L..90D} of the upper limit in the quiet corona, $2.2\times10^5$ km above the photosphere and $T=3.9\times10^6$ K.
We used Baumbach-Allen formula for the density at high altitudes (Figure \ref{density-corona-VALC-C07.eps}). The C7 model terminates at $10^5$ km. Therefore, we added (as lower boundary for the algorithm) at 1 AU representative values of density and temperature for quiet Sun ($T=10^5$ K and $H=10 cm^{-3}$). The source of the emission at the frequencies under study comes from the chromosphere region (see Section \ref{sec4}).


Our model includes three opacity sources: Classic Bremsstrahlung \citep{1979ApJS...40....1K}, Neutral Interaction \citep{1996ASSL..204.....Z,1988A&A...193..189J} and Inverse Bremsstrahlung \citep{1980ZhTFi..50R1847G}. A study of the contribution in the emission for each opacity source in the center of the solar disk can be found in \cite{2011ApJ...737....1D}.

\section{Computations}
PakalMPI \citep{2010ApJS..188..437D} solves the radiative transfer equation for a set of 3D ray paths (or lines of sight from the Earth to the Sun, Figure \ref{colorcaminoopticoEN.eps}). We intersect two geometric systems: i) the spherical heliocentric and ii) the vanishing point from the Earth to the Sun. The intersection of both geometry systems define the spatial points where the radiative transfer equation is solved iteratively in a further step. Each ray path provides the solution for a single pixel in a 2D image. The resolution of the image is fixed by the number of lines of sight in the ray path set.
As the radial atmospheric stratification (density, temperature, etc) is interpolated onto the ray path, then we used the physical conditions from the 1D radial chromospheric models directly.
The center of the Sun is the geometrical point. We control the integration steps ($dz$) on the $z$-axis. The origin of the $z$-axis is the center of the Sun and increments towards the Earth direction. The $y$-axis is perpendicular to $z$-axis and represents the radial distance in the solar disk. We defined $$r = y - R_{sun}$$ as the altitude above the limb, where $R_{sun}$ is the optical solar radii ($6.96\times10^5$ km).

A line of sight 
is obtained from the 2D image using three elements: the pixel resolution, the spatial resolution, and the distance between the source and the image. We 
computed a synthetic spectrum from radio to infrared wavelengths for each pixel in our image.

Additionaly, PakalMPI take 6 parameters to configure the numerical integration: i) the begin (z\_begin), ii) the end (z\_end) of the ray path integration over the $z$-axis, iii) the image resolution in pixels (-r), iv) the spatial resolution or zoom (-Rt), v) the integration step in km (dz), and vi) the minimal parameter to be considered for this algorithm (-min).
In this paper, we used the following configuration for PakalMPI:
\begin{verbatim}
z begin (-z_begin) = -6.96e5 km
z end (-z_end) = 6.96e5 km
Image Resolution (-r) = 9733x9733 px
Spatial Resolution (-Rt) =7.3e5 km
dz (-dz) = 1 km
Minimal Parameter (-min) = 1e-40 [I]
\end{verbatim}

In Table \ref{tbl:a} we defined the frequency and spatial configuration of the simulated spectra. The first column show range in pixels on the $y$-axis, where the pixel 0 is the center of the solar disk. The second column define the range in km and the third column the step in km between each point. The fourth column is the range of frequencies for each spatial point and the last column the step in frequency.
%

\section{Results}\label{sec4}
Figure \ref{limb-brightenning-VALC.eps} show the limb brightening for the VALC model and in Figure  \ref{limb-brightenning-C07.eps} for the C7 model.
We observed three frequency regions in both plots: i) between 2 GHz and 70 GHz where the $T_{\mathrm{b}}$ remains almost constant with height (vertical lines with the same color), ii) a plateau at $2200$ km (VALC) and $2100$ km (C7) above the limb and between 70 GHZ and 1000 GHz where $T_{\mathrm{b}}$ changes abruptly with height (vertical lines change color at these altitudes), and iii) a gradual decrease in $T_{\mathrm{b}}$ for frequencies greater than 1000 GHz (vertical color changes slowly). Between 5 GHz and 100 GHz, the VALC model shows higher limb brightness temperatures than the C7 model. 

The maximum relative limb brightening is computed by taking $T_{\mathrm{b}}(\nu)$ at each position and frequency and normalizing it by the brightness temperature in the center of the solar disk $T_{\mathrm{b0}}(\nu)$ (Figure \ref{plot-nu-vs-Tb0.eps}).
A detailed study of this spectrum in the center of the solar disk can be found in \cite{2013ApJ...762...84D}.

Figures \ref{relative-limb-brightenning-VALC.eps} and \ref{relative-limb-brightenning-C07.eps}
shows the relative limb brightening ($T_{\mathrm{b}}(\nu)/T_{\mathrm{b0}}(\nu)$) for VALC and C7 models respectively. 
C7 show higher relative brightness temperatures in a higher region than VALC model. 
In the case of VALC results, we found an unexpected strong relative limb brightening between 15 and 150 GHz as shown in Figure \ref{relative-limb-brightenning-VALC.eps} (between the vertical white lines) which
is not observed in the absolute limb brightening. 
We also found that at around 40 GHz the extension of this unexpected limb brightening is a maximum. 

Exploring this 
relative maximum in the limb brightening,
we compared the two semi-empirical models 
at 15 GHz and 40 GHz.
In Figure \ref{plot-h-over-the-photosphere-vs-Tb-limb-brightenning.eps} we plot $T_{\mathrm{b}}(\nu)$ vs height above the limb at 15 GHz and 40 GHz for the two models. We found that for the C7 model the limb brightening at 15 GHz is higher than 40 GHz.
  However, we found the opposite in the relative limb brightening ($T_{\mathrm{b}}(\nu)/T_{\mathrm{b0}}(\nu)$) for the VALC model (Figure \ref{plot-h-over-the-photosphere-vs-Tb-relative-limb-brightenning.eps}).


We analyzed the pixel indicating $900$ km above the limb at 15 and 40 GHz (white points in the Figure \ref{relative-limb-brightenning-VALC.eps}). For the VALC model we found a strong decrease at 15 GHz, and a strong increase at 40 GHz in the shape of the relative limb brightening.

Figure \ref{plot-compare-15vs40GHz-Tb.eps} and \ref{plot-compare-15vs40GHz-relative.eps} shows the convergence of $T_{\mathrm{b}}$ and $T_{\mathrm{b}}/T_{\mathrm{b0}}$ on the $z$-axis for VALC and C7 models taking into account two frequencies: 15 and 40 GHz. These figures shows the local contribution to the final  $T_{\mathrm{b}}$ and $T_{\mathrm{b}}/T_{\mathrm{b0}}$. For both cases, the main contribution to the emission process takes place around  $42\times10^3$ and $46\times10^3$ km on the $z$-axis. Note for the VALC at 15 GHz a plateau is formed between $43\times10^3$ and $45\times10^3$ Km. This structure is not observed at 40 GHz.




Figure \ref{plot-compare-15vs40GHz-Tau-total-sum.eps} show the optical depth for Bremsstrahlung and H${}^{-}$ on the z-axis for the pixel indicating $900$ km above the limb. VALC model found a region (on the z-axis) at $43\times10^3$ where the opacity decreases slowly until $46\times10^3$ km (compared with C7 model).

Figure \ref{plot-compare-15vs40GHz-z-vs-r.eps} show that for $43\times10^3$ km and $46\times10^3$ km on z-axis corresponds the distance above the photosphere (r) of $2.1\times10^3$ km and $2.3\times 10^3$ km respectively, i.e. we found that the plateau in the radiative transfer process is originated by the plateau in the radial temperature profile of the VALC model. We applied this principle for each altitude in this study. Figure \ref{plot-compare-15vs40GHz-r-vs-Tr-VALC.eps} shows the temperature limits to be evaluated by PakalMPI for a particular height above the limb, the yellow boxes defines the height above the limb (ray path) and the arrow points the temperature limit on the temperature model.

With figures \ref{relative-limb-brightenning-VALC.eps} and \ref{relative-limb-brightenning-C07.eps}, we computed the solar radii at frequency $\nu$ as the isophote where $$T_{\mathrm{b}}(\nu)/T_{\mathrm{b0}}(\nu) = 0.5.$$
In order to compare the synthetic spectrums, we included high resolution observations of the solar radii with the solar eclipse occultation technique \citep{1994IAUS..154..139C} carried out by the James Clerk Maxwell Telescope. The computed solar radii from observations using polinomial function degree 2 is:
\begin{equation}
  R_{sun} (\nu[GHz]) [km] = 3.42\times10^4 - 1.81\times10^4\log(\nu) + 2420\log^2(\nu)
\end{equation}
The Figure \ref{solar-radii-VALC-C07-obs.eps} compares the computed theoretical solar radii versus the high resolution observations. The theoretical radii from both models is close in all the frequency ranges.
In Figure \ref{solar-radii-VALC-C07-obs-error.eps}, we ploted the diferences between the synthetic solar radii and the observed. We found a maximum difference of $12000$ km at 20 GHz.

\section{Conclusions}
\begin{figure}
   \centerline{
\includegraphics[width=1.0\textwidth]{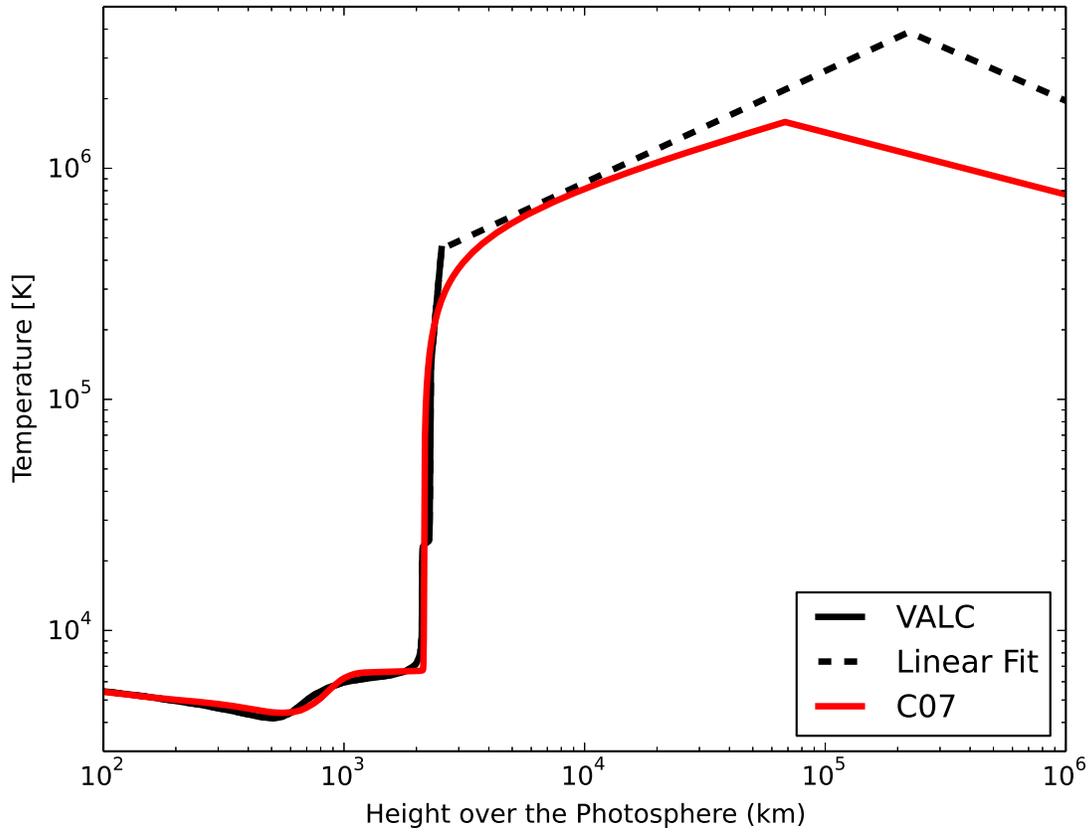}}
\caption{Radial temperature profile for VALC (continuous black) and C7 (continuous red). The dashed line is the linear interpolation to fit corona altitudes for VALC model. The C7 model includes the ambipolar diffusion that generates a profile without plateau at $20000$ km above the photosphere.}
\label{temperature-corona-VALC-C07.eps}
\end{figure}

\begin{figure}
   \centerline{
\includegraphics[width=1.0\textwidth]{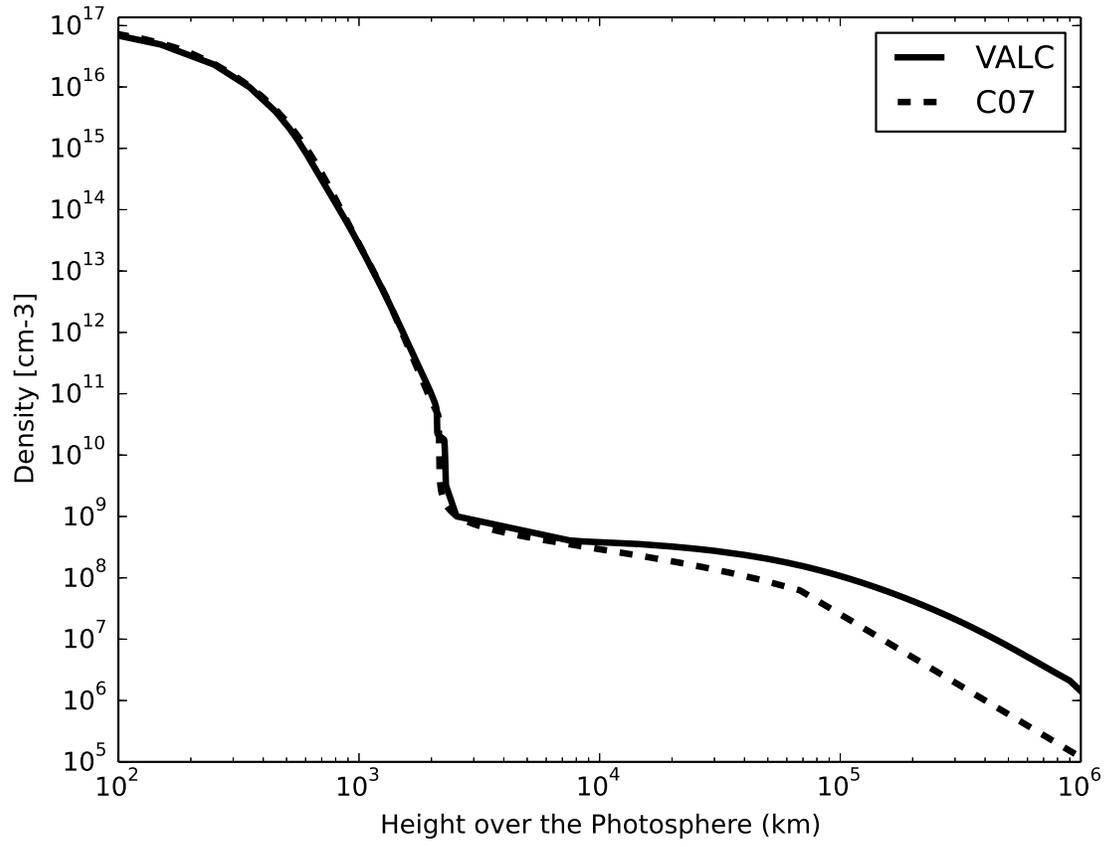}}
\caption{Density models for the corona. In black VALC + Baumbach-Allen formula and dashed line for C7 model.}
\label{density-corona-VALC-C07.eps}
\end{figure}

\begin{figure}
   \centerline{
\includegraphics[width=1.0\textwidth]{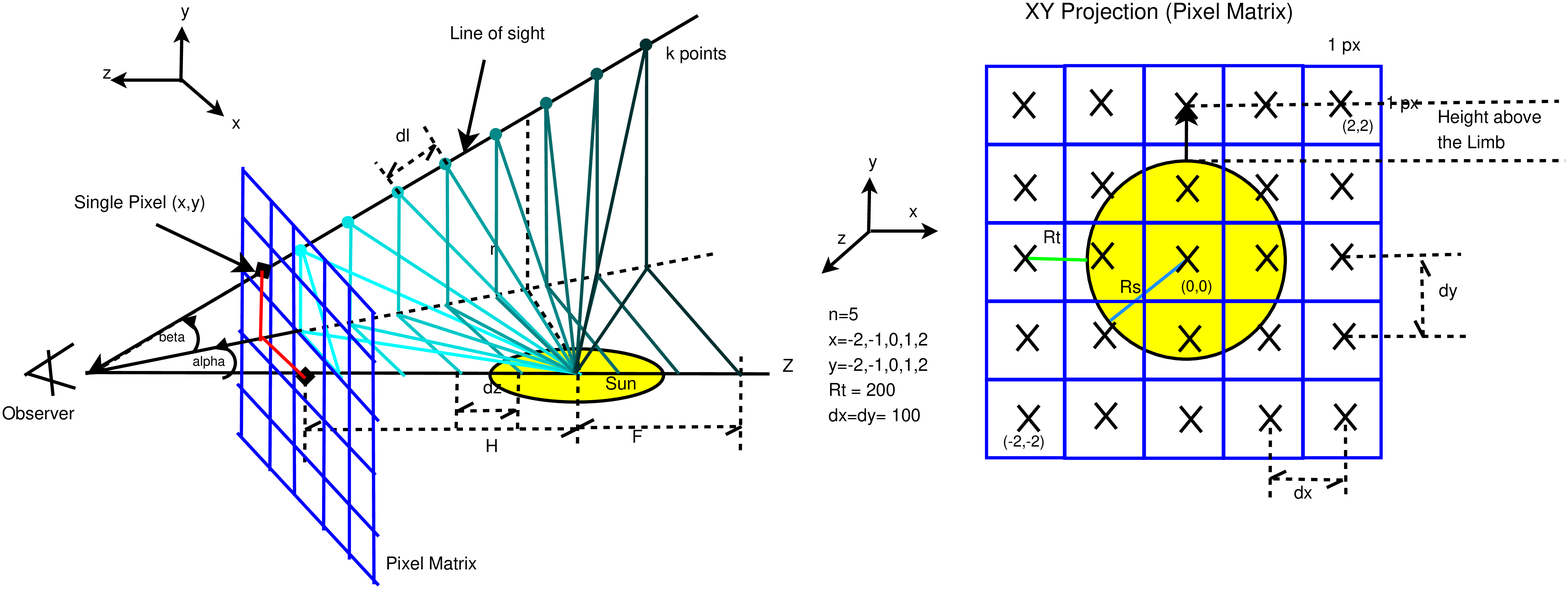}} 
\caption{Left panel: 3D geometry used to solve the radiative transfer equation. Each line of sight (or ray path) defines the pixel matrix. We can solve this geometry using 2 parameters:  the size of the image in km (Rt) and the resolution in pixels of the Pixel Matrix (n). Right panel: Example of Pixel Matrix using $n=5$ and $Rt=200$ km. In this figure we define the height above the limb as the distance between the border of solar disk projected in the Pixel Matrix and a point in the Matrix.}
\label{colorcaminoopticoEN.eps}
\end{figure}

We presented the first set of normalized solar synthetic spectrums from 2 GHz to 10 THz between $0$ km and $4500$ km above the solar limb. We used both semi-empirical models: VALC \citep{1981ApJS...45..635V} and C7 \citep{2008ApJS..175..229A} as input in the physical conditions of the chromosphere (Figures \ref{temperature-corona-VALC-C07.eps} and \ref{density-corona-VALC-C07.eps}). The radiative transfer equation was solved by PakalMPI \citep{2010ApJS..188..437D}
 in a 3D geometry above the limb (Figure \ref{colorcaminoopticoEN.eps}). NLTE computations were carried out for calculate ionization states, including hydrogen, electron density, and H${}^-$. Bremsstrahlung, H${}^-$, and Inverse Bremsstrahlung \citep{2011ApJ...737....1D} where used in the computation of the optical depth.

\begin{figure}
   \centerline{
\includegraphics[width=1.0\textwidth]{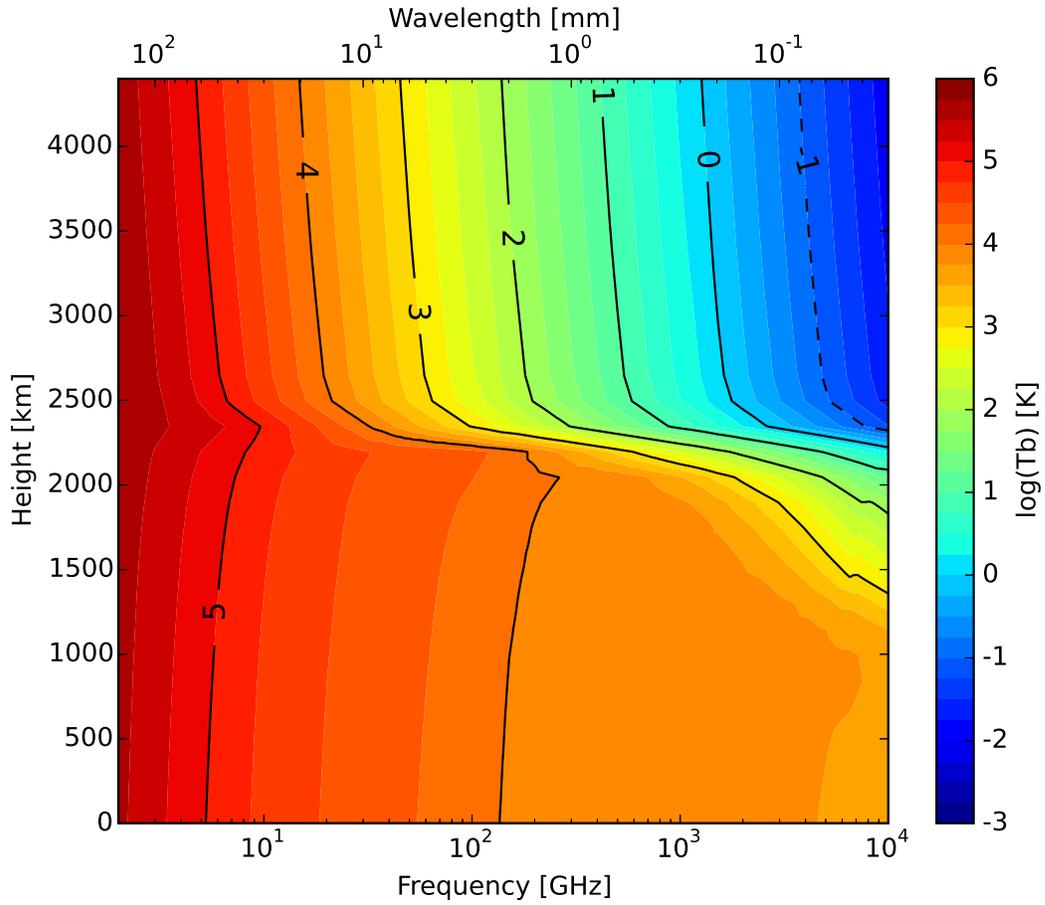}}  
\caption{Limb brightening using VALC model: x-axis is the frequency in GHz, y-axis is the height above the solar limb (off limb), and in colors we plot the Brightness Temperature ($\log(T_{\mathrm{b}})$). The figure shows a region at $2200$ km above the limb and between 50 and 1000 GHz where $T_{\mathrm{b}}$ strongly decrease.}
\label{limb-brightenning-VALC.eps}
\end{figure}

\begin{figure}
   \centerline{
\includegraphics[width=1.0\textwidth]{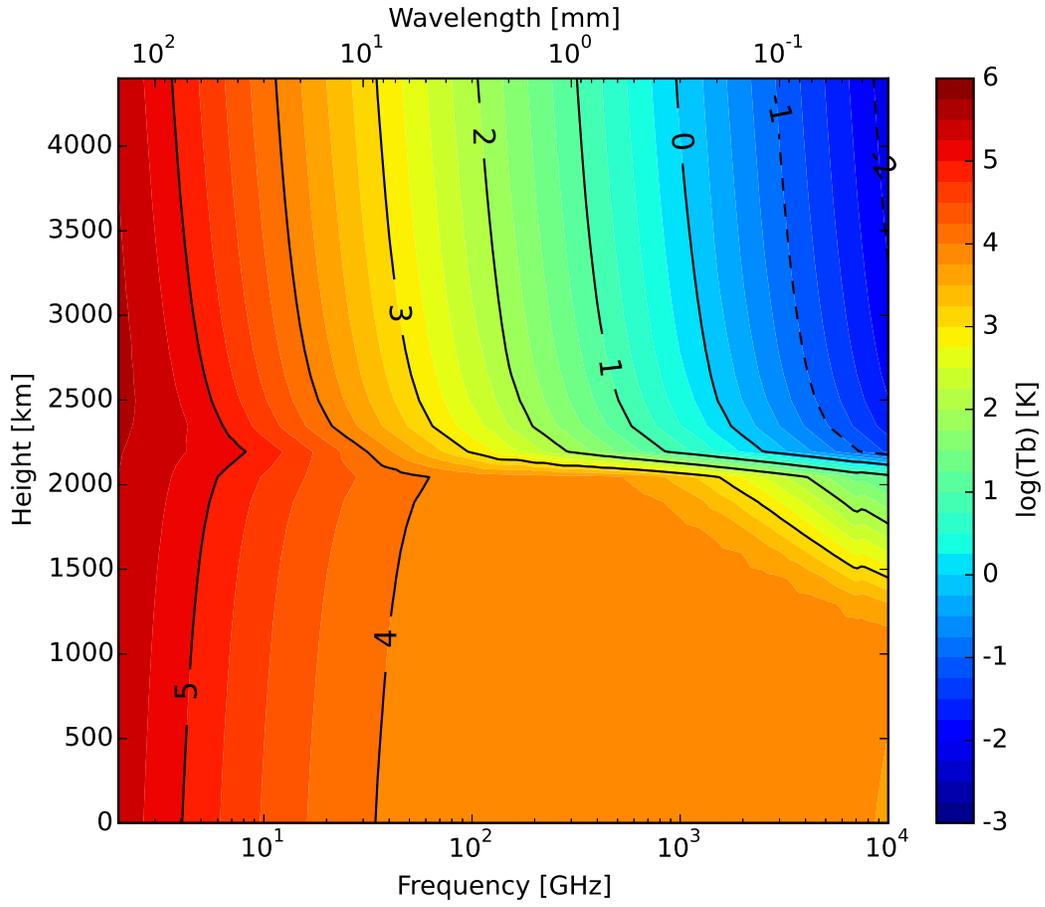}} 
\caption{Limb brightening using C7 model. This model shows lower $T_{\mathrm{b}}$ compared to VALC model but we also observe the interface in $T_{\mathrm{b}}$ between 50 and 1000 GHz at $2100$ km above the limb.}
\label{limb-brightenning-C07.eps}
\end{figure}

\begin{figure}
   \centerline{
\includegraphics[width=1.0\textwidth]{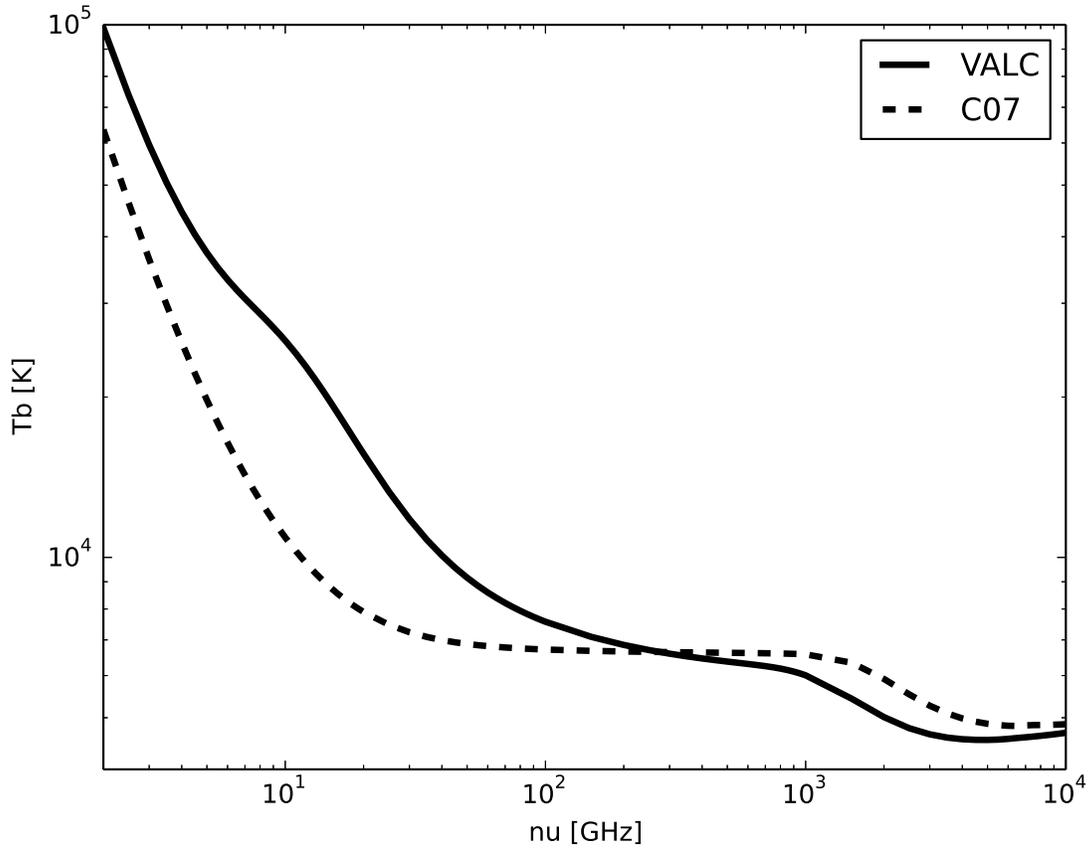}} 
   \caption{Synthetic spectrum in the center of the solar disk for VALC (continuous) and C7 (dashed). VALC shows higher $T_{\mathrm{b}}$ than C7 for frequencies lower than 300 GHz. In frequencies higher than 300 GHz VALC shows lower $T_{\mathrm{b}}$ than C7 model. A detailed study of the spectrum in the center of the solar disk can be found in \cite{2011ApJ...737....1D}}
\label{plot-nu-vs-Tb0.eps}
\end{figure}

\begin{figure}
   \centerline{
\includegraphics[width=1.0\textwidth]{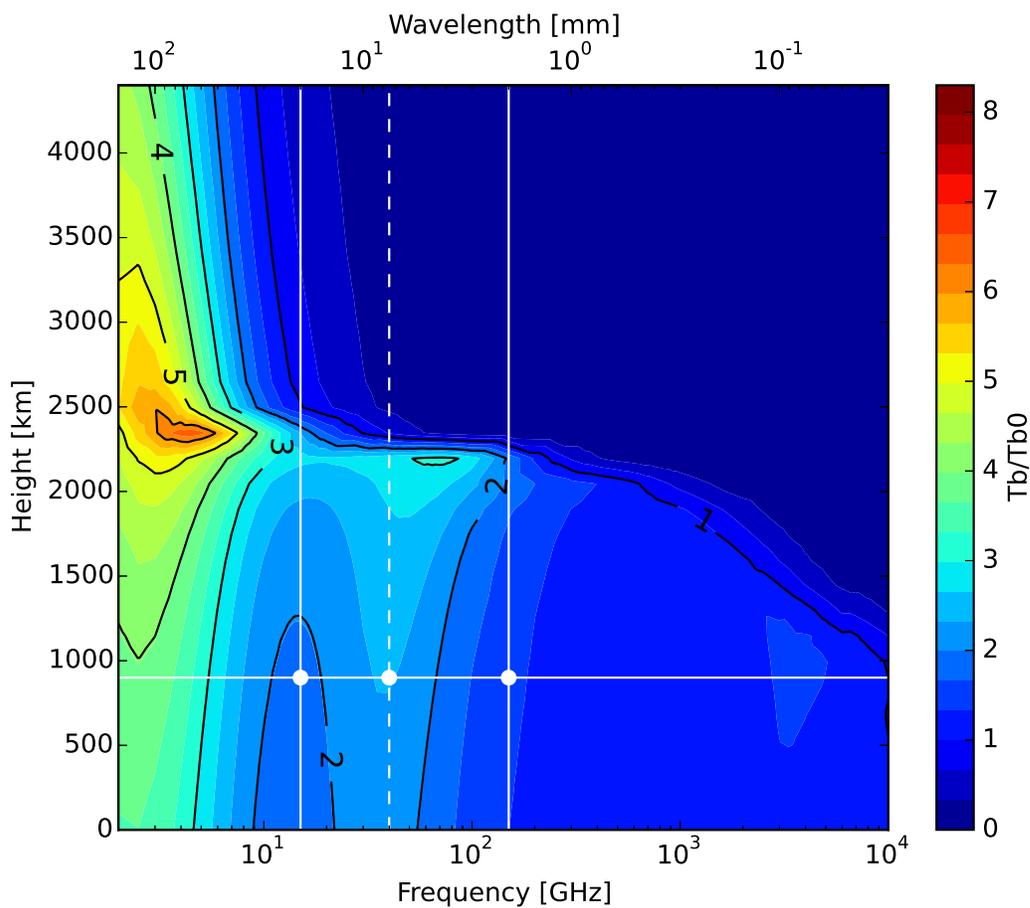}} 
\caption{Relative limb brightening using VALC model. In colors we plot $T_{\mathrm{b}}/T_{\mathrm{b0}}$ for each frequency for each height above the limb using the spatial configuration defined in Table 1. The figure shows an unexpected high relative limb brightening between 15 and 150 GHz. The dashed vertical line shows the frequency where the unexpected relative limb brightening has a maximum extension (40 GHz).}
\label{relative-limb-brightenning-VALC.eps}
\end{figure}

\begin{figure}
   \centerline{
\includegraphics[width=1.0\textwidth]{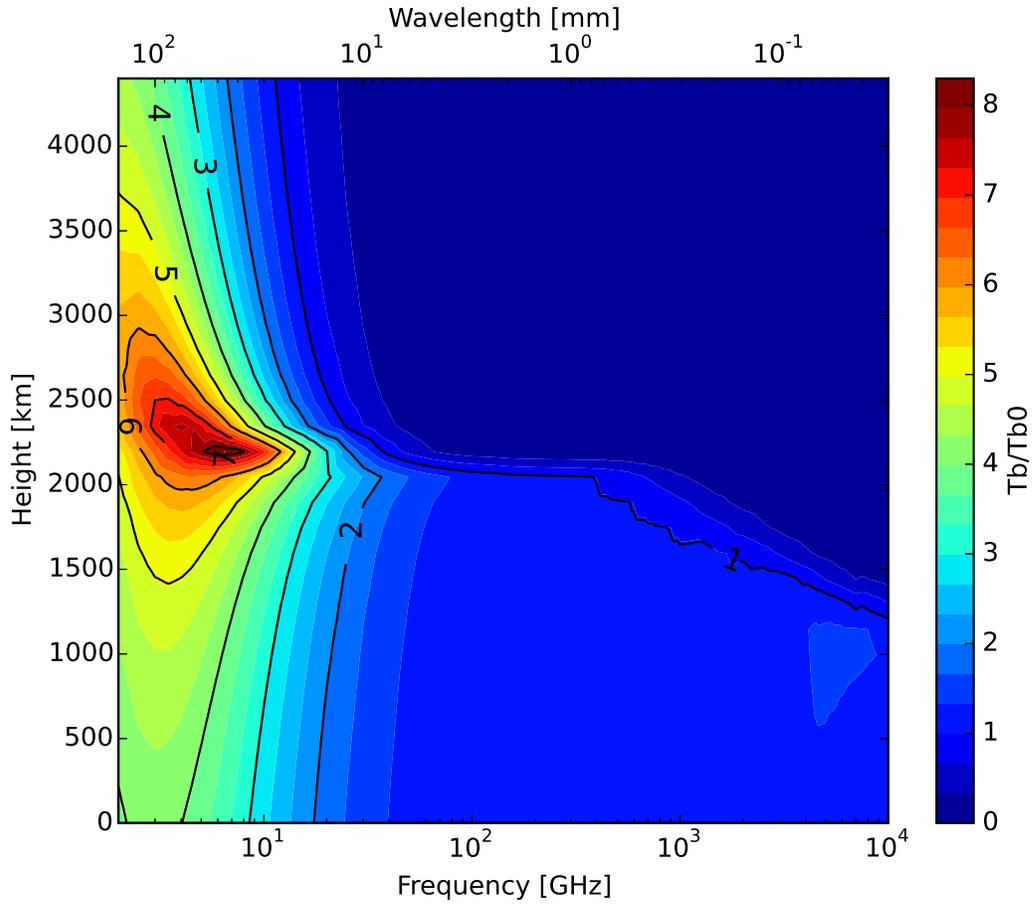}} 
\caption{Relative limb brightening using C7 model. This model shows higher relative brightness temperatures than VALC model. The peak of maximum relative limb brightening is between 2 and 10 GHz.}
\label{relative-limb-brightenning-C07.eps}
\end{figure}

\begin{figure}
   \centerline{
\includegraphics[width=1.0\textwidth]{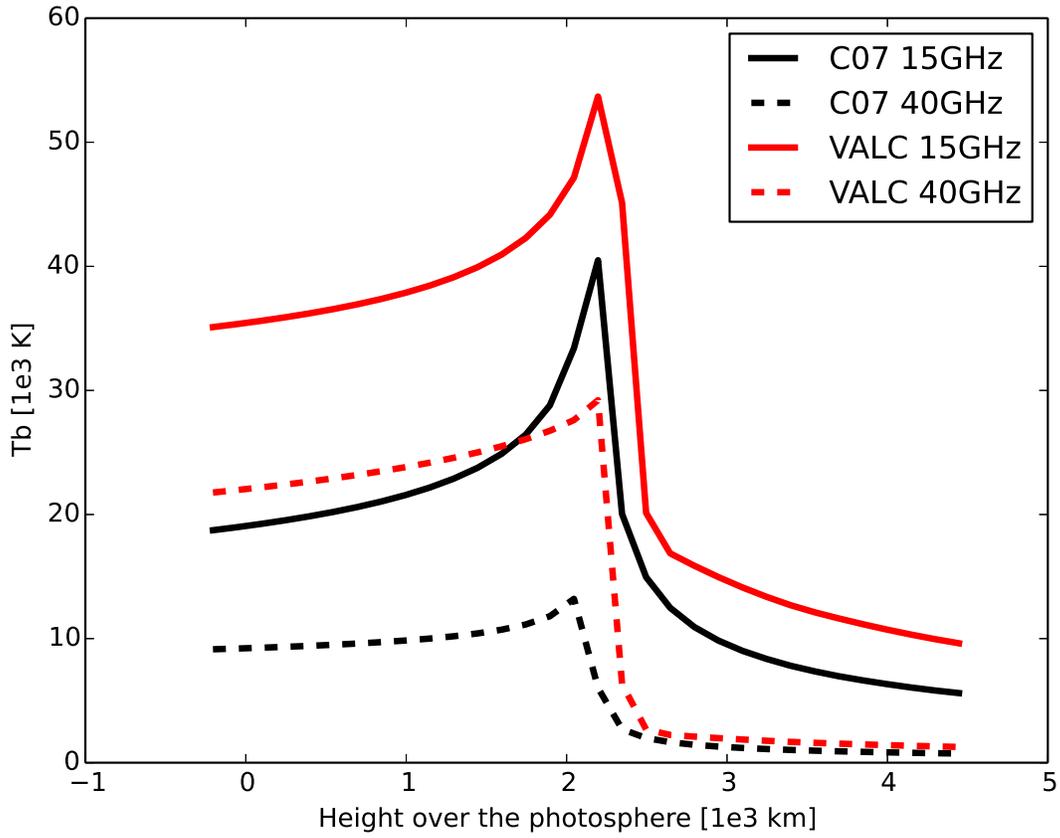}}
\caption{Limb brightening for 15 GHz and 40 GHz using C7 and VALC models. C7 shows lower $T_{\mathrm{b}}$ than VALC model. Both models show the classical limb brightening.}
\label{plot-h-over-the-photosphere-vs-Tb-limb-brightenning.eps}
\end{figure}

\begin{figure}
   \centerline{
\includegraphics[width=1.0\textwidth]{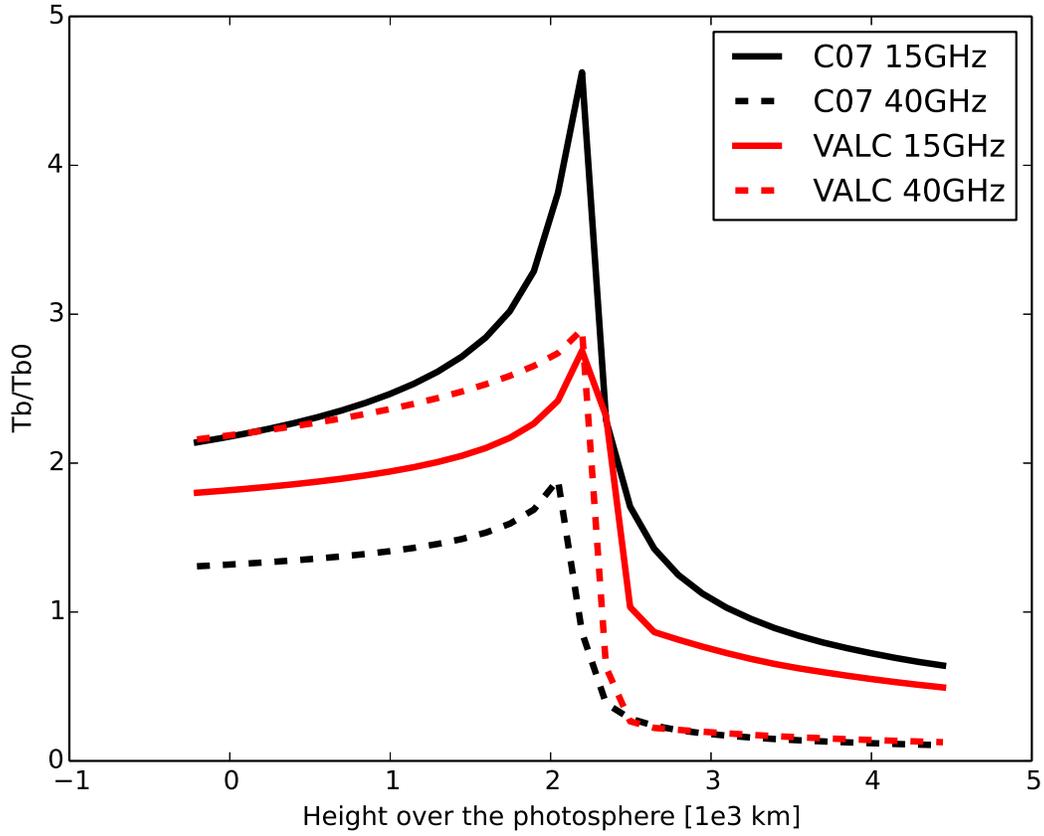}}
   \caption{Comparison between relative limb brightening for 15 and 40 GHz using both models at solar limb altitudes. C7 shows the expected result (higher frequency has lower $T_{\mathrm{b}}$) like classical limb brightening. VALC shows the opposite. The maximum limb brightening at 15 GHz and 40 GHz
   has the same $T_{\mathrm{b}}$.}
\label{plot-h-over-the-photosphere-vs-Tb-relative-limb-brightenning.eps}
\end{figure}

\begin{figure}
   \centerline{
\includegraphics[width=1.0\textwidth]{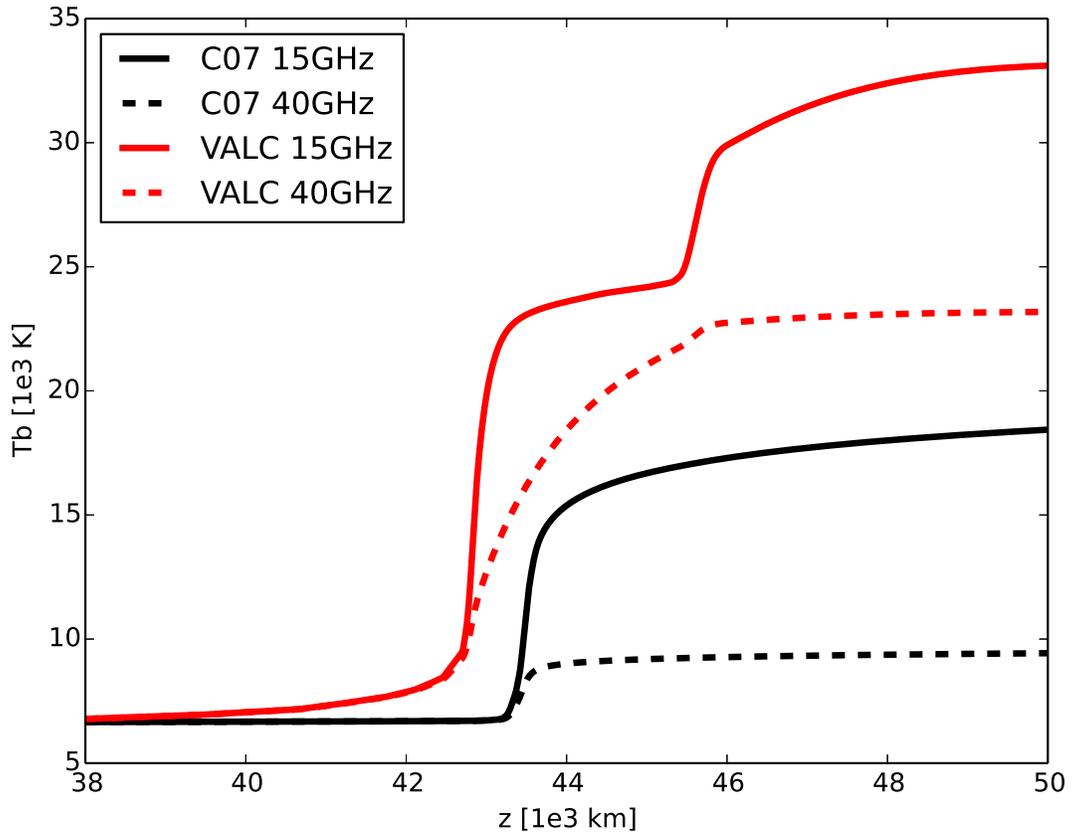}}
   \caption{$T_{\mathrm{b}}$ for 15 GHz and 40 GHz for VALC and C7 models in the ray path projected onto the z axis at a height above the limb of $900$ km. The profile for VALC at 15 GHz presents two major increases in the emission at $42500$ and $45500$ km in $z$.}
\label{plot-compare-15vs40GHz-Tb.eps}
\end{figure}

\begin{figure}
   \centerline{
\includegraphics[width=1.0\textwidth]{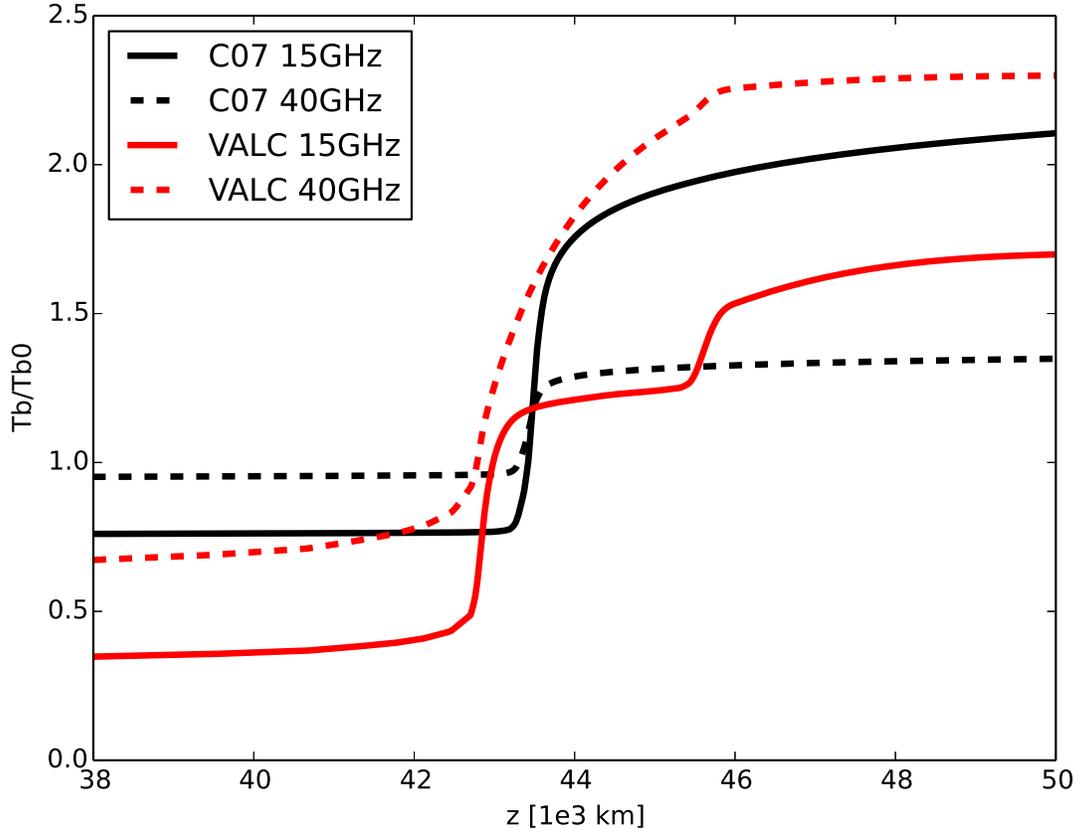}}
\caption{When dividing $T_{\mathrm{b}}$ by $T_{\mathrm{b0}}$, the large difference in $T_{\mathrm{b0}}$ computed by VALC model at 15 GHz and 40 GHz results in a high relative brightness temperature at 40 GHz than 15 GHz. We also observe the second increase in relative limb brightening at $45500$ km in $z$ that is not shown in the other profiles.}
\label{plot-compare-15vs40GHz-relative.eps}

\end{figure}

\begin{figure}
   \centerline{
\includegraphics[width=1.0\textwidth]{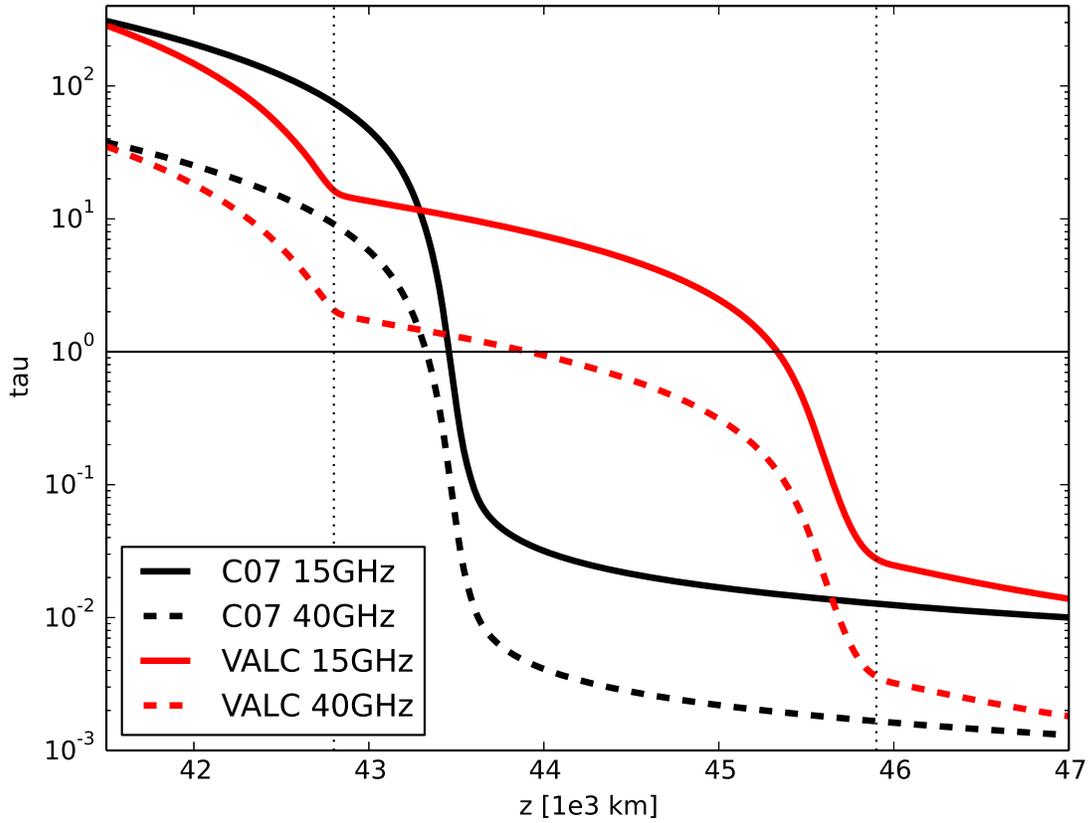}}
\caption{Optical depth for C7 (black) and VALC (red) models at 15 GHz (continuous) and 40 GHz (dashed) at altitude above the limb at $900$ km. We found that the cause of the second region of emission for VALC at 15 GHz is the plateau in tau close to 1 is between $43000$ and $46000$ km in $z$.}
\label{plot-compare-15vs40GHz-Tau-total-sum.eps}
\end{figure}

\begin{figure}
   \centerline{
\includegraphics[width=1.0\textwidth]{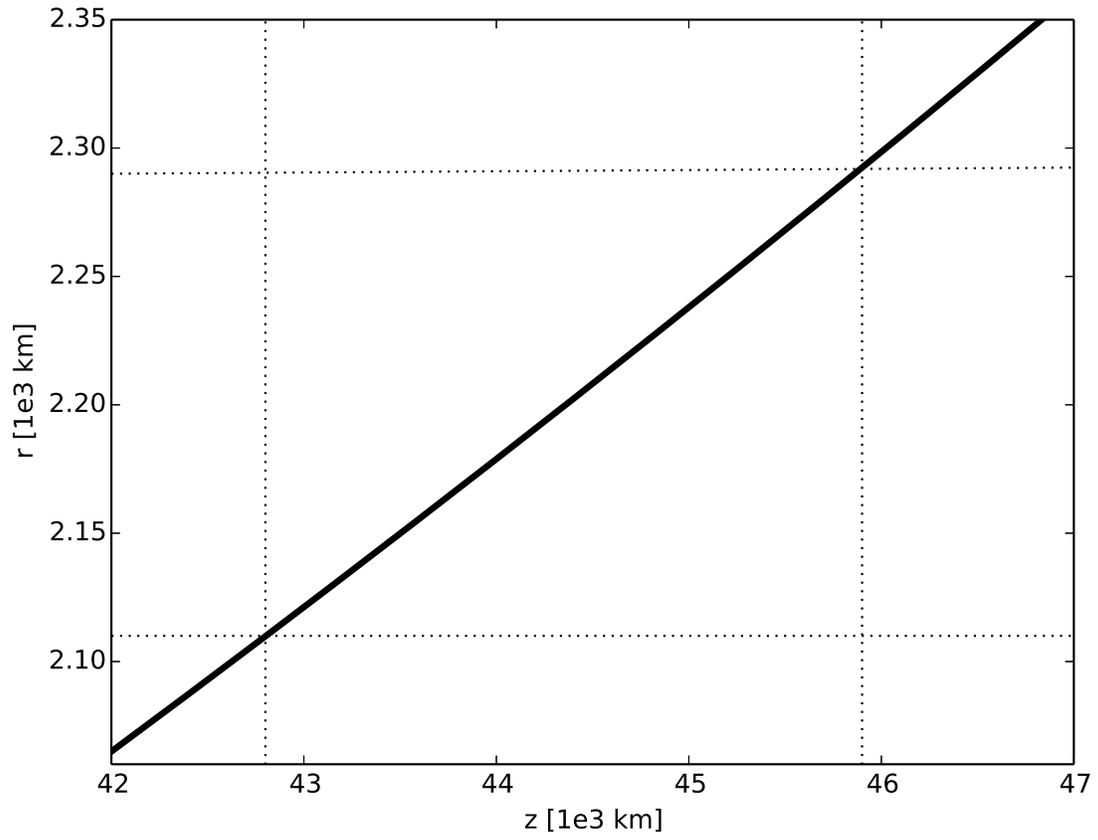}}
\caption{Photosphere distance r in function of z for a ray path at 900 km above the limb.}
\label{plot-compare-15vs40GHz-z-vs-r.eps}
\end{figure}

\begin{figure}
   \centerline{
\includegraphics[width=1.0\textwidth]{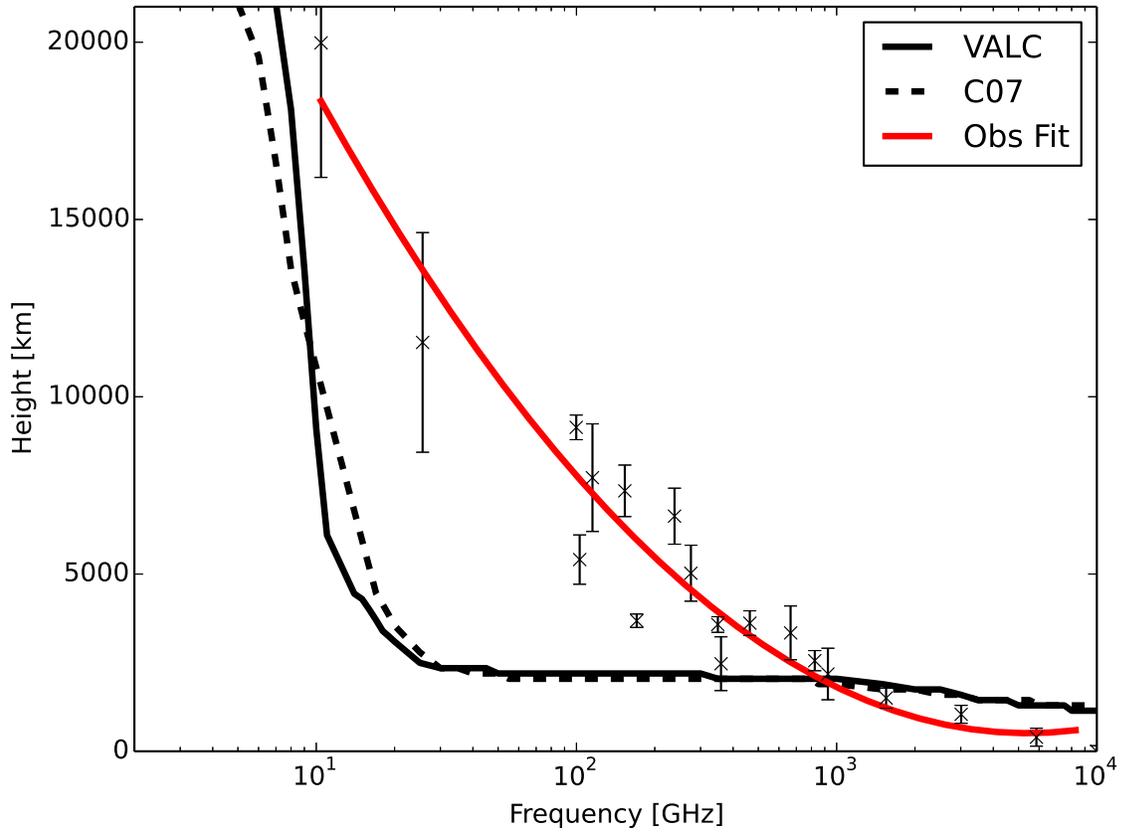}}
\caption{Simulated solar radii for VALC (continuous blackline) and C7 (dashed black line) models vs observations. The cross points with error bars are the observations from \cite{1994IAUS..154..139C} and the continuos red line is their polinomial fit. The emission from the chromospheric models can not reproduce the solar radii.}
\label{solar-radii-VALC-C07-obs.eps}
\end{figure}

\begin{figure}
   \centerline{
\includegraphics[width=1.0\textwidth]{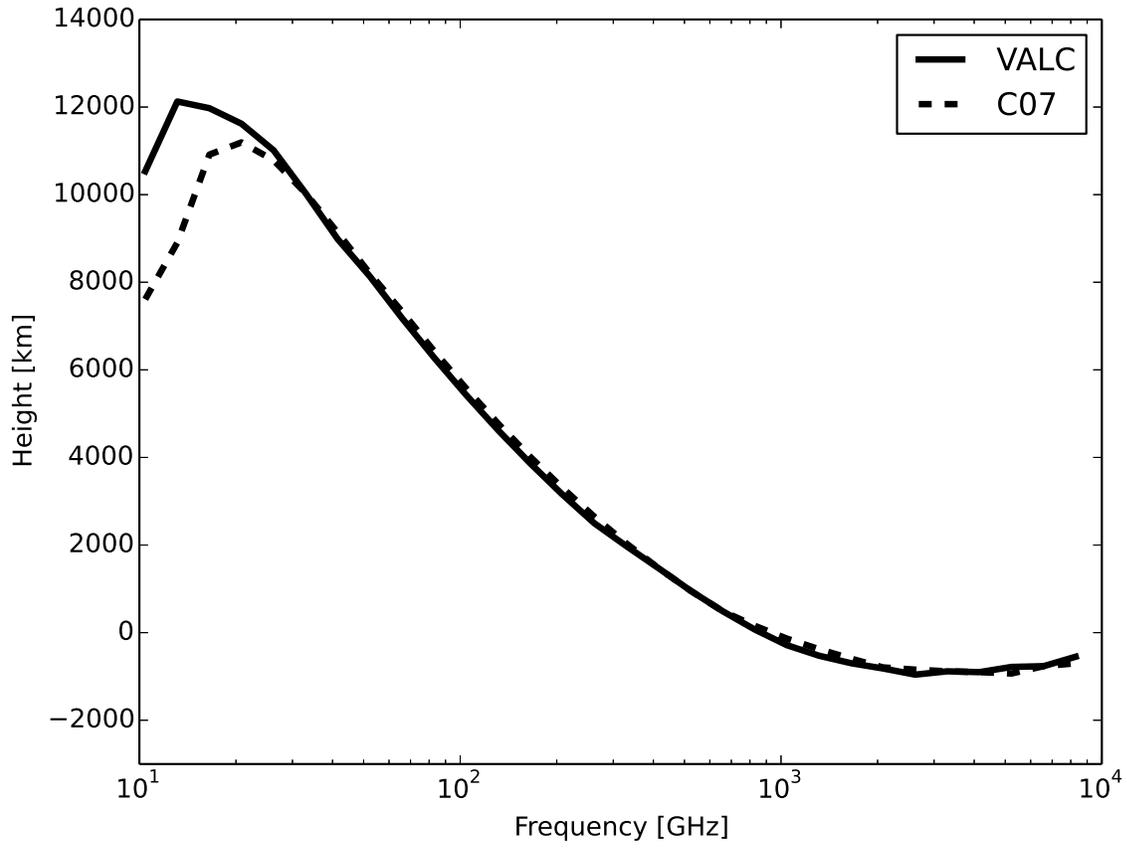}}
\caption{Difference between observations from \cite{1994IAUS..154..139C}  and this work. The continuos line is the difference for the VALC model and the dashed line is for the C07 model. The maximum difference for both model is at 20 GHz.}
\label{solar-radii-VALC-C07-obs-error.eps}
\end{figure}

 The synthetic spectrums computed by PakalMPI in the paths above the limb were taked into account to calculate the theoretical solar limb brightening (Figures \ref{limb-brightenning-VALC.eps} and \ref{limb-brightenning-C07.eps}). The theoretical spectrum in the center of the solar disk (Figure \ref{plot-nu-vs-Tb0.eps}) was used to compute the relative limb brightening $T_{\mathrm{b}}/T_{\mathrm{b0}}$ at altitudes above the limb (Figures \ref{relative-limb-brightenning-VALC.eps} and \ref{relative-limb-brightenning-C07.eps}). The radiative transfer process  between 15 and 150 GHz (where we found an unexpected strong relative limb brightening) was analyzed (Figures
 \ref{plot-h-over-the-photosphere-vs-Tb-limb-brightenning.eps},
\ref{plot-h-over-the-photosphere-vs-Tb-relative-limb-brightenning.eps},
\ref{plot-compare-15vs40GHz-Tb.eps},
\ref{plot-compare-15vs40GHz-relative.eps},
\ref{plot-compare-15vs40GHz-Tau-total-sum.eps}, and
\ref{plot-compare-15vs40GHz-z-vs-r.eps}).
Finally, we calculated the theoretical solar radii and compared the results (Figures \ref{solar-radii-VALC-C07-obs.eps} and
\ref{solar-radii-VALC-C07-obs-error.eps}) with previous observations by  \cite{1994IAUS..154..139C}.

\begin{figure}
   \centerline{
\includegraphics[width=1.0\textwidth]{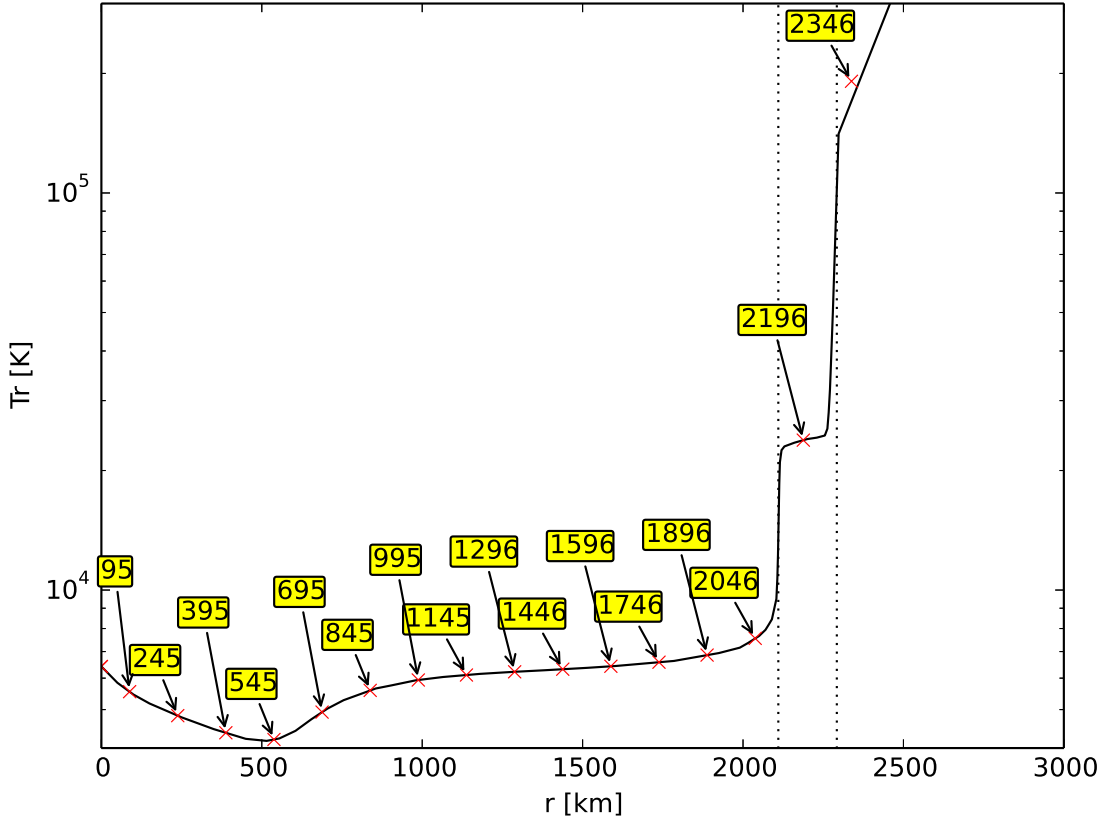}}
   \caption{We show 
     the radial temperature profile $T_r$ from the VALC model that is used for computing the $T_{\mathrm{b}}$. In yellow are the heights in  km above the limb where the radiative transfer equation is computed. The red cruxes show the lower boundary in temperature on the radial profile that is used to solve the RTE for each ray path. For example at $2196$ km above the limb the model used temperatures from the radial temperature profile above $2200$ km.}
\label{plot-compare-15vs40GHz-r-vs-Tr-VALC.eps}
\end{figure}

In both semi-empirical models, we found three regions in frequency where the rise of $T_{\mathrm{b}}$ with respect to the height above the limb changes significantly: 2 GHz - 70 GHz, 70 GHz - 1000 GHz, and 1000 GHz - 10000 GHz. VALC model shows higher limb brightness temperatures than the C7 model between 5 GHz and 100 GHz. The relation between the limb brightening and the radial temperature profile is related with the minimal distance between the ray path (used to compute the radiative transfer equation) and the photosphere (Figure \ref{plot-compare-15vs40GHz-r-vs-Tr-VALC.eps}). The minimal distance is the lower boundary in the radial temperature profile used to compute the emission.


The normalization in frequency of the spectrum by their emission in the center of the solar disk (Figures \ref{relative-limb-brightenning-VALC.eps} and \ref{relative-limb-brightenning-C07.eps}) provides a tool to test semi-empirical models of the chromosphere above the limb. We found that the unexpected relative limb brightening in the VALC model between 15 and 150 GHz is caused by the plateau in temperature of the radial temperature profile of the VALC model (Figure \ref{plot-compare-15vs40GHz-Tau-total-sum.eps}). The inclusion of ambipolar diffusion (C7 model) reduces the
relative limb brightening
in this region (Figures \ref{plot-compare-15vs40GHz-Tb.eps} and \ref{plot-compare-15vs40GHz-relative.eps}).

The solar radii obtained by the relative limb brightening show differences of $12000$ km at $20$ GHz, $5000$ km at $100$ GHz, and $1000$ km at $3$ THz if there are compared against previous observations \citep{1994IAUS..154..139C}. These diferences are observed for both semi-empirical models.

Our results show that the stratified chromospheric models used in this study are not enough to reproduce the solar radii at frequencies lower than $400$ GHz, and suggest other structures evolved in the emission process (Figure \ref{solar-radii-VALC-C07-obs.eps}).


In this context, recent observations of emission of spicules-off the solar limb at H$\alpha$ \citep{2014ApJ...795L..23S} show a clear emission of $4000$ km above the limb, indicating that the individual spicules structures extend above $4000$ km. This kind of spicules are called type II (from quiet Sun regions). 
The highest altitudes measured of the individual spicules rise $8000$ km above the limb. However, observations at 20 GHz \citep{1994IAUS..154..139C} have shown the existence of 
structures that rise $20000$ km above the limb. There is a maximum difference between the computed theoretical solar radii and observations by \cite{1994IAUS..154..139C} of $12000$ km at 20 GHz (see Figure \ref{solar-radii-VALC-C07-obs-error.eps}) that chromospheric semi-empirical models \citep{1981ApJS...45..635V,2008ApJS..175..229A} can not reproduce.
 This H$\alpha$ emission observed from the spicules type II \citep{{2014ApJ...795L..23S}} suggest that for altitudes lower than $8000$ km above the limb, temperature of the region is lower than the ionized temperature for hydrogen. However, this temperature difference is not enough to explain the solar radii at low frequencies. 
At high frequencies, we found that the difference between simulations and observations decreases to $5000$ km at 100 GHz and $1000$ km at 3 THz. 

If the spicules are the cause of the observed large solar radii at low frequencies, then the spicules should rise to altitudes of about $20000$ km above the limb. The millimeter emission at $8000$ km above the solar limb can be explained by Bremsstrahlung of full ionized hydrogen which we can not observe in H$\alpha$.

Our results, suggest that the inclusion of the micro-structure in the high chromosphere must be considered when investigating the solar limb brightening and therefore in the models focused in the center of the solar disk as shown in previous works \citep{1981SoPh...69..273A,1991ApJ...381..288R,1994IAUS..154..139C, 2007PASJ...59S.655D,2015ApJ...804...48I,2006A&A...456..697W,2015A&A...575A..15L}.

Finally, this study is usefull to undestand the limb occultation by solar microwave sources in eruptive events \citep{1995hola} and to characterize planetary transits for solar-like stars at millimeter, sub-millimeter and infrared wavelengths. For the case of planetary transits, the values of radii and their limb brightening are fundamental in the characterization of the light curve \citep{2013ApJ...777L..34S} recorded during planet transits at these wavelengths.

\acknowledgments
This work was supported by the C\'atedras CONACyT Fellowship. We are grateful to J. Americo Gonzalez-Esparza and Ernesto Aguilar for their valuable comments.






%
%

\bibliographystyle{apj}


\bibliography{libros}

\clearpage



\begin{table}
\caption{Spacial configuration of the simulations. The first column defines range in pixels on the $y$-axis (where the pixel 0 is the center of the solar disk), the second column the range in km, the third column the step in km between each point. The fourth column is the range of frequencies for each spatial point and the last column the step in frequency.}\label{tbl:a}
\begin{tabular}{  c c c c c }
\multicolumn{3}{ c }{$y$-axis} &  \multicolumn{2}{ c }{Frequencies [GHz]} \\
Range [px]&Range [Km]&Step [km]&Range&Step\\
\hline
\multirow{4}{*}{[4638,4669]}     & \multirow{4}{*}{[695700,700350]}  & \multirow{4}{*}{150} &  [2,10]   & 0.5  \\
                                 &   &                       & [10,100]  & 5  \\
                                 &   &                       & [100,1000] & 50  \\
                                 &   &                       & [1000,10000] & 500  \\
\hline
\multirow{1}{*}{[4680,4780]}     & [702000,717000]  & \multirow{1}{*}{1500} &  [2,11]   & 1  \\
\hline
\end{tabular}
\end{table}

\clearpage

\clearpage

\clearpage

\clearpage

\clearpage

\clearpage

\clearpage

\clearpage

\clearpage

\clearpage

\clearpage

\clearpage

\clearpage

%
%
\clearpage

\clearpage

\clearpage

\clearpage

\clearpage

\end{document}